\documentstyle[twoside,fleqn,espcrc2]{article}

\newcommand{\AmS}{{\protect\the\textfont2
  A\kern-.1667em\lower.5ex\hbox{M}\kern-.125emS}}

\title{Electroweak Strings, Zero Modes and Baryon Number
}

\author{Tanmay Vachaspati\address{Isaac Newton Institute,
        University of Cambridge, \\
        20 Clarkson Road, Cambridge, CB3 0EH, United Kingdom.}%
        \thanks{Address from January 1: Department of Physics,
                Case Western Reserve University,
                Cleveland, OH 44106, U.S.A.}
       }

\begin{document}

\begin{abstract}
The Dirac equations for leptons and quarks in the background
of an electroweak $Z-$string have zero mode solutions. If
two loops of electroweak string are linked, the zero modes
on one of the loops interacts with the other loop via an
Aharanov-Bohm interaction. The effects of this interaction are
briefly discussed and it is shown that the fermions induce
a baryon number on linked loops of $Z-$string.
\end{abstract}

\maketitle

\section{Introduction}

Topological defects have received widespread attention in
the particle physics community since the discovery of
string \cite{hnpo} and monopole \cite{thp} solutions in the
70's in certain toy field theories. These discoveries have
also provided impetus to considerable work in cosmology and
astrophysics since the early universe is precisely the arena
in which defects are likely to play a major role.

Over the last two decades \cite{yn,nm,tv,mb2tv}, it has been
gradually recognised that the standard model of the
electroweak interactions also contains classical solutions
that closely resemble topological defects. There are two
kinds of strings in the standard model - the
$W-$ and $Z-$ strings. Furthermore, the $Z-$string terminates on
magnetic monopoles. This recognition immediately makes the
entire two decades long study of topological defects relevant
to the world of particle physics and opens up some surprises too.

One of the recent surprises is that
linked loops of $Z-$string carry baryon number \cite{tvgf}.
This can be shown by a simple exercise. The baryon
number current satisfies the well-known anomaly equation
which can be integrated to yield the change in the
baryon number between two different times.
For field configurations in which all the gauge
fields except for the $Z$ gauge field are zero, we get
\begin{equation}
\Delta B =  N_F {{\alpha^2} \over {16\pi^2}} cos(2\theta_W )
\Delta
\int d^3 x ~{\vec Z} \cdot ( {\vec \nabla} \times {\vec Z} )
\end{equation}
where, $\alpha = \sqrt{g^2 + g'^2}$ and $N_F$ is the number of
families. When two loops of string
that are linked once decay into a final vacuum state
with zero baryon number, the integration can be done and yields
the baryon number of the initial string configuration:
\begin{equation}
B = 2N_F cos(2\theta_W ) \ .
\label{1}
\end{equation}

This calculation of the baryon number of two linked loops is
an indirect calculation since no direct reference has been made
to the fermions that actually carry the baryon number. (Indirectly,
the fermions have come in because the anomalous baryon current
conservation equation has included the effects of the fermions.)
In recent and ongoing work with Jaume Garriga, we have directly
considered the effects of leptons and quarks in the singly linked
string background. In this way we have recovered eqn. (\ref{1}) and also
found some other surprises. These are:

(i) The ground state of the fermions in the background of singly
linked string loops is lower than that if the loops are not linked.

(ii) In the ground state, the strings carry non-vanishing
electromagnetic currents but zero net charge.

Our treatment of this problem enables us to evaluate any other
charge of the linked string configuration that we may wish to
calculate though some quantities will require a numerical evaluation.
I now outline the calculation; details may be found in our
paper \cite{jgtv}.

\section{Fermions on linked strings}

There are three basic facts that need to be put together to
understand how the leptons and quarks contribute to say the
baryon number of linked strings. The first fact is that
the fermions interact non-trivially with the strings and have
zero energy states in the background of the string. These
zero modes were first discovered by Jackiw and Rebbi \cite{rjcr}
and constructed for the $Z-$string by Earnshaw and
Perkins \cite{mewp}. The second fact is that the leptons and
quarks in the standard model have an Aharanov-Bohm interaction
with the $Z-$string. As a result of this interaction, the
zero modes on linked loops have a dispersion relation which is
different from the dispersion relation for zero modes on
unlinked loops. The third fact is that
an adiabatic change in the dispersion relation can lead to the
production of particles and hence to the production of certain
charges. Alternately, the ground states of the two configurations
(linked and unlinked strings) can differ in their energy and
charges. The change in the fermion energy eigenvalues as the
background configuration is changed is known as ``spectral
flow'' and has been studied in a variety of circumstances. The
early work most relevant for us is the work on Schwinger
electrodynamics by Manton \cite{nmse}.

In the background
of a $Z-$string, the zero modes satisfy the dispersion relation:
\begin{equation}
E = \pm p
\end{equation}
where, the electron and $d$ quark travel in one direction (- sign)
and the neutrino and $u$ quark in the opposite direction (+ sign).
(Similarly we can deal with the other fermion families.)
If two loops are linked, the dispersion  relation is modified
to
\begin{equation}
E = \pm (p - q Z)
\end{equation}
where, $q$ is the $Z-$charge on the lepton or quark in units of the
charge on the Higgs field and $Z = n/a$ where, $n$ is the number of
strings threading a loop and $a$ is the radius of the loop which
we assume to be circular and much larger than the thickness of the
string. (In the following we choose units such that $a=1$. Then,
$Z=n$.) Given this dispersion relation, we can find the energy of the
Dirac sea. Naively, the energy is infinite but we can use
zeta function regularization to find the energy contribution
due to one fermion
\begin{equation}
E = -{1 \over {24}} + {1 \over 2} [p_F - qn \pm {1 \over {2}} ]^2
\label{2}
\end{equation}
where $p_F$ labels the Fermi momentum (highest filled state) and
is an integer and the sign depends on the fermion in question.
The expression for $E$ contains the Casimir energy
contribution since the loop is closed. This is the same for linked or
unlinked strings and does not interest us. In the second term,
the $qn$ piece is due to the linkage. If the loops are unlinked,
this term is absent and the ground state energy for unlinked loops
is
\begin{equation}
E[n=0] = -{1 \over {24}} + {1 \over {8}} = {1 \over {12}} \ .
\end{equation}
As a function of $qn$, this energy is a maximum. The minimum
energy can be attained if $qn$ is half integral because then
we can fill the states (that is, choose $p_F$) such that
the second term in (\ref{2}) is zero and the energy
$E = -1/24$. For the leptons and quarks occurring
in the standard model, $q$ is not an integer and so the ground
state energy of linked strings is lower than that of unlinked
strings. But since the charges are related to $sin^2\theta_W$,
they are probably irrational and the energy functional can never
take on its minimum possible value.

We can evaluate the charges on the linked or unlinked loops
in a similar manner. Each energy eigenstate that is filled carries
a certain charge and adding up the charges in the Dirac sea
means that we have to find
\begin{equation}
Q = c \sum_{p = -\infty}^{p_F - qZ} 1
\end{equation}
where each fermion carries a charge $c$.
Once again this sum is divergent. But we can make sense of the sum
by regularizing it using zeta function regularization. We
then have:
\begin{equation}
Q = c \sum_{p = -\infty}^{p_F - qZ} (p-qZ)^0
              = \pm c[ p_F - qZ \pm  {1\over 2} ] \
\end{equation}
where the signs need to be chosen according to the fermion whose
charge we are evaluating.
Note that we have chosen to write $1 = (p-qZ)^0$ rather than
$1=p^0$ since $p-qZ$ is the gauge invariant combination. (This
valuable trick was used by Manton in Ref. \cite{nmse}.)

Now we can sum over all fermions with the appropriate charges
$q$ and $c$ when the Fermi levels $p_F$ are evaluated by minimizing
the energy. We also have to choose the states to be colour singlets.
When we put all this arithmetic together we find the electric charge
\begin{equation}
Q_A = 0
\end{equation}
and the baryonic charge
\begin{equation}
Q_B = 2 N_F cos(2\theta_W ) \ .
\end{equation}
This agrees with the indirect calculation described in the
introduction.

\section{Conclusions}

The linking of loops of string is detected by the fermions that
live on the strings even though the loops of string themselves
are well separated. This is possible due to the Aharanov-Bohm
interaction of the fermions with the strings and leads to non-trivial
charges and currents on linked strings.

It is known that in general string knots are characterized by an
infinite sequence of numbers and the link invariant is only the first
of this sequence. What we have found is that the fermions living on
the strings are sensitive to this first knot invariant.
It would be amusing if a situation could be devised in which the fermions
not only experience the linkage but also get affected by the higher knot
invariants.

\end{document}